\documentclass[12pt,preprint]{aastex}

\def  \be   {\begin{equation}}
\def  \ee   {\end{equation}}
\def  \beq  {\begin{eqnarray}}
\def  \eeq  {\end{eqnarray}}

\shorttitle{Growth of magnetic field in helical turbulence}
\shortauthors{Malyshkin \& Boldyrev}

\begin{document}

\title{Magnetic dynamo action in helical turbulence}

\author{Leonid Malyshkin\altaffilmark{1} \& Stanislav Boldyrev\altaffilmark{2}}
\affil{${~}^1$Department of Astronomy \& Astrophysics, 
University of Chicago, 5640 S. Ellis Ave., Chicago, IL 60637; {\sf leonmal@uchicago.edu}} 
\affil{${~}^2$Department of Physics, University of Wisconsin-Madison, 1150 University Ave., 
Madison, WI 53706; {\sf boldyrev@wisc.edu}}


\begin{abstract}
We investigate magnetic field amplification in a turbulent velocity field with 
nonzero helicity, in the framework of the kinematic Kazantsev-Kraichnan model. 
We present the numerical solution of the model for the practically important 
case of Kolmogorov 
distribution of velocity fluctuations, with a large magnetic Reynolds 
number. We find that in contrast to the nonhelical case where growing magnetic 
fields are described by a few bound eigenmodes concentrated inside the inertial interval of 
the velocity field,  
in the helical case the number of bound eigenmodes considerably increases; moreover, new 
unbound eigenmodes appear. Both bound and unbound eigenmodes contribute to the 
large-scale magnetic field. This indicates a limited applicability of the 
conventional alpha model of a large-scale dynamo action, which captures only unbound 
modes.              
\end{abstract}


\section{Introduction}
Magnetic fields in planets and stars, protogalaxies and galaxies, and possibly 
intergalactic medium are generated 
due to random stretching of magnetic field lines by turbulent motion of highly 
conducting fluids or plasmas in which these lines are frozen 
\citep[e.g.,][]{lynden-bell,parker,moffatt78,kulsrud1,zweibel,schekochihin,kulsrud2}. 
It is natural to expect that such a dynamo mechanism can amplify magnetic fields at 
scales smaller than the correlation scales of the velocity fields.  
However, magnetic fields observed in 
astrophysical systems often appear to be correlated at larger scales. Such magnetic 
fields can be explained if one assumes that the velocity field ${\bf v}({\bf x, t})$ 
possesses nonzero kinetic 
helicity, that is,  $H=\int {\bf v}\cdot ({\nabla \times {\bf v}})\, d^3 x\neq 0$.  
To describe the large-scale fields, one generally applies the 
alpha model~\citep{skr,moffatt78,kulsrud1}. 
This model is obtained if one averages the induction equation
\begin{eqnarray}
\partial_t {\bf B}=\nabla \times ({\bf v}\times {\bf B})+\eta \nabla^2 {\bf B},
\label{induction}
\end{eqnarray}
where $\eta$ is resistivity or magnetic diffusivity,  over small-scale 
fluctuations of the velocity and magnetic fields, assuming that these fluctuations are 
much weaker and concentrated at the scales much smaller than the scales of the growing 
large-scale field. As a result one obtains the alpha model equation for the large-scale 
magnetic field $\overline{\bf B}({\bf x}, t)$, that is, the magnetic field 
averaged over the scales larger than the scales of velocity fluctuations, 
\begin{eqnarray}     
\partial_t \overline{\bf B}=\alpha \nabla \times \overline{\bf B}+\beta \nabla^2 \overline{\bf B}.  
\label{alpha}
\end{eqnarray} 
In this equation, $\alpha\sim \overline{{\bf v}\cdot ({\nabla \times {\bf v}})} \tau_0$, 
and $\beta \sim v_0 l_0$, where $v_0$ is a characteristic velocity, $l_0$ is the 
characteristic scale, and $\tau_0\sim l_0/v_0$ is a characteristic eddy turnover time (decorrelation 
time) of fluid fluctuations. In astrophysical systems resistivity is very small, so that turbulent 
velocity field effectively generates small-scale magnetic fluctuations, which grow much 
faster than the large-scale field. In this case, the applicability of the mean-field 
equation~(\ref{alpha}) is questionable 
\citep[see, e.g., the discussion in][]{vainshtein1,diamond,blackman}.

To investigate this question, we use a solvable model of kinematic dynamo, introduced by 
Kazantsev~(1968) and Kraichnan~(1968). In this model, velocity is assumed to be Gaussian, 
with zero mean, $\langle{\bf v}\rangle=0$, and the covariance tensor
\beq 
\langle {v^i}({\bf x},t){v^j}({\bf x}',t') \rangle \!=\!
\kappa^{ij}(|{\bf x}-{\bf x}'|)\delta(t-t'), 
\label{V_V_TENSOR}
\eeq
where $\kappa^{ij}$ is an isotropic tensor,
\beq
\kappa^{ij}({\bf x})\!=\!\kappa_N 
\left(\delta^{ij}-\frac{x^ix^j}{x^2}\right)+
\kappa_L \frac{x^ix^j}{x^2}+g\epsilon^{ijk}x^{k}.  
\label{KAPPA}
\eeq
Here $\langle \rangle$ denotes ensemble average, 
$\epsilon^{ijk}$ is the unit anti-symmetric pseudo-tensor and
summation over repeated indices is assumed. 
The first two terms on the right-hand side of (\ref{KAPPA})
represent the mirror-symmetric, nonhelical part, while function $g(x)$
describes the helical part of the velocity fluctuations. 
For an incompressible velocity field (the only case we are considering 
here), we have $\kappa_N(x)=\kappa_L(x)+x\kappa'_L(x)/2$, where the prime 
denotes  derivative with respect to~$x=|{\bf x}|$. Therefore, 
to describe the velocity field, we need to specify only two 
independent functions, say, $\kappa_L(x)$ and $g(x)$. 
The magnetic field correlator can similarly be expressed as
\begin{eqnarray}
\langle B^i({\bf x}, t)B^j(0,t)\rangle = 
M_N\left(\delta^{ij}-\frac{x^ix^j}{x^2}\right) 
+ M_L\frac{x^ix^j}{x^2}+K\epsilon^{ijk}x^k,
\label{B_B_TENSOR}
\end{eqnarray} 
where the corresponding solenoidality constraint implies 
$M_N(x,t)=M_L(x,t)+(x/2)M'_L(x,t)$. To describe the magnetic correlator we 
therefore need only two functions, $M_L(x, t)$ and $K(x, t)$, corresponding to 
magnetic energy and magnetic helicity. 

Suppose that the velocity field~(\ref{V_V_TENSOR}) is given. 
The problem is then to find the correlation function~(\ref{B_B_TENSOR}) of the magnetic field. 
In the nonhelical case, $g(x)\equiv 0$, it was established by \citet{kazantsev68} that the problem 
is reduced to a quantum mechanical problem with imaginary time:
\begin{eqnarray}
-\partial_t \psi ={\hat{\cal H}}\psi.
\label{quantum} 
\end{eqnarray}
More precisely, given the velocity correlator~(\ref{V_V_TENSOR}) 
and magnetic resistivity $\eta$, one constructs the self-adjoint 
Hamiltonian~${\hat{\cal H}}$. The magnetic correlator $M_L(x, t)$ is then mapped to 
the ``wave function''~$\psi(x, t)$. If the equation~(\ref{quantum}) has growing 
solutions $\psi(x, t)$, corresponding to negative eigenvalues of ${\hat{\cal H}}$,  
then dynamo action is possible. One can show that the smaller the resistivity~$\eta$, 
the deeper the effective potential in the quantum mechanical problem~(\ref{quantum}), 
and, therefore, nonhelical dynamo is always possible when the magnetic Reynolds number, 
${\rm Rm}\propto \eta^{-1}$, is large enough~\citep{vainshtein,boldyrev-cattaneo,iskakov}. 
The quantum mechanical representation~(\ref{quantum}) is important since it ensures 
that the eigenvalues are real and the eigenfunctions are mutually orthogonal, which 
allows one to apply a variational principle for estimating dynamo growth rates. 
The Kazantsev-Kraichnan model of nonhelical dynamo action has been well investigated 
in the literature.

The situation is more complex when the velocity field possesses nonzero helicity, 
i.e., $g(x)\neq 0$. In this case, given kinetic energy~$\kappa_L(x)$ and kinetic helicity~$g(x)$, 
one needs to solve {\em two} coupled partial differential equations for functions
$M_L(x,t)$ and $K(x, t)$ related to magnetic energy and magnetic helicity. 
Such equations were first derived by \citet{vainshtein}.  
Due to their complexity, there have been relatively fewer results obtained for the 
helical case  \citep[e.g.,][]{pouquet,kulsrud,kim,brandenburg,blackman,brandenburg2}. 
However, it is the helical case that is practically more important since astrophysical 
systems generally possess nonzero helicity. Moreover, while magnetic fields at small 
(velocity) scales are naturally expected on the base of equation~(\ref{quantum}), 
explanation of astrophysically observed large-scale magnetic fields commonly requires 
helicity effects, as, e.g., in equation~(\ref{alpha}). 

Recently, it has been established in \citet{boldyrev05} that the \citet{vainshtein} 
equations also possess a self-adjoint structure, and can be reduced to a quantum 
mechanical ``spinor'' form:
\begin{eqnarray}
-\partial_t \psi^{\alpha} ={\hat{\cal H}}^{\alpha \beta}\psi^{\beta},
\label{spinor}
\end{eqnarray}
where $\alpha=\{1,2\}$ and summation over repeated indices is assumed. Similarly to 
equation~(\ref{quantum}), the self-adjoint Hamiltonian ${\hat{\cal H}}^{\alpha\beta}$ depends 
on kinetic energy and kinetic helicity, $\kappa_L(x)$ and $g(x)$, and on magnetic resistivity~$\eta$. 
The two components of the $\psi^{\alpha}(x, t)$ function are then related to 
functions $M_L(x,t)$ and $K(x, t)$ in the magnetic correlator~(\ref{B_B_TENSOR}).

Similar structures of the dynamo equations~(\ref{quantum}) and~(\ref{spinor}) allow one to 
investigate them on the same footing. In particular, one can compare the growth rates and the 
eigenvalues associated with helical and nonhelical dynamo action, and address the important 
question of whether the large-scale magnetic field generated due to helical dynamo action 
is described by the alpha model~(\ref{alpha}). This is the goal of the present work. 
We assume that the velocity field 
has the Kolmogorov spectrum and that the Reynolds number and the magnetic Reynolds number are 
large. Then we present the full numerical solution of the helical dynamo model~(\ref{spinor}), 
i.e., we find its eigenvalues and eigenfunctions. We demonstrate that at least at the kinematic 
stage, the alpha model~(\ref{alpha}) may provide a nonadequate description of large-scale 
fields, since it misses the rapidly growing large-scale eigenmodes. In the next 
section we present our main results; a detailed discussion will be presented elsewhere. 
nonhelical
\section{Numerical solution of the helical dynamo model}

The self-adjoint equations for functions $M_L(x, t)$ and $K(x, t)$
were derived in \citet{boldyrev05}. We will be interested in eigenmodes of these equations, 
and therefore assume that both functions depend on time as $\exp(\lambda t)$. It is 
then convenient to introduce the auxiliary functions $w_2(x)$ and $w_3(x)$ defined as:
\beq
M_L=\frac{\sqrt{2}e^{\lambda t}}{x^2}w_2(x), 
\quad
K=-\frac{e^{\lambda t}}{\sqrt{2}x^4}
\left[x^2 w_3(x)\right]'.
\label{M_AND_K}
\end{eqnarray}
The eigenmode equation that we solve then takes the form:
\begin{eqnarray}
\left[\begin{array}{cc}
-\frac{\sqrt{2}}{x}{\hat E}\frac{\sqrt{2}}{x}-\lambda &
 \frac{\sqrt{2}}{x^3}C\frac{d}{dx} x^2\\
-{x^2}\frac{d}{dx}C\frac{\sqrt{2}}{x^3} &
 x^2\frac{d}{dx}\frac{B}{x^4}\frac{d}{dx}{x^2}-\lambda
\end{array}
\right]
\left[\!\!\begin{array}{c}
w_2\\
w_3
 \end{array}\!\!\right]=0,
\label{W2_AND_W3}
\eeq
where
\beq
\begin{array}{lcl}
{\hat E}&=&-\frac{1}{2}x\frac{d}{dx}B\frac{d}{dx}x 
+ \frac{1}{\sqrt{2}}(A-xA'),\\
A(x)&=&\sqrt{2}\left[2\eta+\kappa_N(0)-\kappa_N(x)\right],\\
B(x)&=&2\eta+\kappa_L(0)-\kappa_L(x),\\
C(x)&=&\sqrt{2}\left[g(0)-g(x)\right]x,
\end{array}
\label{EABC}
\eeq
and primes denote derivatives with respect to $x$.  
Equations~(\ref{M_AND_K})--(\ref{EABC}) describe the growth of the
magnetic field in the Kazantsev-Kraichnan
model. Equation~(\ref{W2_AND_W3}) is self-adjoint, which
guarantees that all growth rates $\lambda$ are real.
The system~(\ref{W2_AND_W3}) cannot be solved analytically for  
general velocity correlation functions~$\kappa_L(x)$ and~$g(x)$. 
Below we solve equation~(\ref{W2_AND_W3}) 
numerically, and concentrate on functions $w_2(x)$ and $w_3(x)$ 
since they define the magnetic field correlator uniquely and contain 
all the information about magnetic energy and magnetic helicity. 

It is useful to note the Fourier transformed version of 
Equation~(\ref{KAPPA}): 
\beq
\kappa^{ij}({\bf k})=F(k)\left(\delta^{ij}-\frac{k^ik^j}{k^2}\right)
+iG(k)\epsilon^{ijl}k^l,
\label{KAPPA_FOURIER}
\eeq
where functions $F(k)$ and $G(k)$ can be expressed in terms of the
three-dimensional Fourier transforms of $\kappa_L(x)$ and
$g(x)$~\citep{monin71}, and 
\beq
\langle B^i({\bf k},t)B^{*j}({\bf k},t)\rangle =
F_B(k,t)\left(\delta^{ij}-\frac{k^ik^j}{k^2}\right) 
- i\frac{H_B(k,t)}{2k^2}\epsilon^{ijl}k^l,
\label{MAGNETIC_SPECTRA}
\eeq
where $F_B(k,t)$ is the magnetic energy spectral function, 
$\langle|{\bf B}({\bf k},t)|^2\rangle=2F_B(k,t)$, and $H_B(k,t)$
is the spectral function of the electric current helicity,
$\langle {B^i}^*({\bf k},t)\: i\epsilon^{ijl}k^jB^l({\bf k},t)\rangle=H_B(k,t)$.

First, as easily derived from (\ref{W2_AND_W3}) the asymptotic behavior 
of functions $w_2(x)$ and $w_3(x)$ when $x\ll\sqrt{\eta/\kappa_L''(0)}$ is 
\beq
w_2=x^2+O(x^4), 
\qquad
w_3=\xi x^3+O(x^5),
\label{W2_W3_SMALL_X}
\eeq
where, without loss of generality, we use scaling $w_2/x^2\to 1$ as
$x\to 0$, and coefficient $\xi$ is a free parameter, related to
the averaged helicity of the electric current,  
$\langle{\bf B}\cdot({\bf\nabla}\times{\bf B})\rangle=6K(0)=-15\sqrt{2}\xi$.

Second, the asymptotic behavior of functions $w_2(x)$ and $w_3(x)$
as $x\to\infty$ depends on the magnetic field growth rate
$\lambda$. If $\lambda>\lambda_0\equiv g^2(0)/[\kappa_L(0)+2\eta]$, then
as $x\to\infty$ the asymptotic eigenfunctions are
\beq
\left[\!\!\begin{array}{c}
w_2\\
w_3
\end{array}\!\!\right]\propto e^{-k_rx}
\left[\!\!\begin{array}{c}
\cos(k_ix+\phi)\\
\sin(k_ix+\phi)
\end{array}\!\!\right],
\label{W_LARGE_X_TRAPPED}
\eeq
where $k_r=\sqrt{\lambda-\lambda_0}/\sqrt{\kappa_L(0)+2\eta}$
and $k_i=\sqrt{\lambda_0}/\sqrt{\kappa_L(0)+2\eta}$.
For $\lambda\le\lambda_0$, the asymptotic behavior of 
eigenfunctions $w_2(x)$ and $w_3(x)$ at $x\to \infty$ becomes a mixture of cosine and
sine standing waves, $\cos(k_i^\pm x+\phi^\pm)$ and 
$\sin(k_i^\pm x+\phi^\pm)$, with wavenumbers
$k_i^\pm=(\sqrt{\lambda_0}\pm\sqrt{\lambda_0-\lambda})/\sqrt{\kappa_L(0)+2\eta}$.
This asymptotic behavior suggests an analogy between
equation~(\ref{W2_AND_W3}) and quantum mechanics. The
spatially localized (bound) eigenfunctions~(\ref{W_LARGE_X_TRAPPED}) with
$\lambda>\lambda_0$ correspond to ``particles'' trapped by the
potential provided by velocity fluctuations, while the nonlocalized (unbound)
eigenfunctions with $\lambda\le\lambda_0$ correspond to ``traveling
particles.'' Eigenvalue $\lambda=\lambda_0$ corresponds to the 
fastest growing unbound eigenmode.

To find the eigenmodes, we proceed as follows. For given
values of $\lambda$ and $\xi$ we integrate ordinary
differential equations~(\ref{W2_AND_W3}) numerically by the fourth-order
Runge-Kutta method. Due to quite disparate scales present in the problem,  
we use a nonuniform numerical grid with 
the grid steps chosen as $\Delta x(x)\propto B^{1/2}(x)$ for $x\ge\sqrt{\eta}$ 
and $\Delta x(x)\propto(x/\sqrt{\eta})B^{1/2}(\sqrt{\eta})$ for $x<\sqrt{\eta}$.
The integration is done starting at a small value of 
$x=x_{\rm min}\ll\sqrt{\eta/\kappa_L''(0)}$, where the asymptotic 
result~(\ref{W2_W3_SMALL_X}) holds. We integrate up to a large value of 
$x=x_{\rm max}$ that is chosen in such way that the numerical solution 
is still stable, while it has already reached its asymptotic behavior 
for large values of $x$.\footnote{
Our results do not depend on the exact choice of the boundaries of 
the computational interval $x_{\rm min}\le x\le x_{\rm max}$.
}
By matching the numerically calculated values $w_2(x_{\rm max})$ and
$w_3(x_{\rm max})$ to their analytic asymptotic solutions at
$x=x_{\rm max}$, we find the eigenvalues (the magnetic field
growth rates) $\lambda$ and coefficients $\xi$.  It turns out that
these eigenvalues are discrete for the bound (localized) eigenmodes, 
i.e., $\lambda=\lambda_n>\lambda_0$ and $\xi=\xi_n$ ($n=1,2,3...$). 
In contrast, the eigenvalues of the unbound (nonlocalized) modes 
are continuous; they exist for any choice
of $\lambda\le\lambda_0$ and for any realizable value 
of $\xi$.\footnote{Given $M_L(x,t)$, function $K(x,t)$ cannot be
chosen arbitrarily, its Fourier image must satisfy the realizability condition
$|H_B(k,t)|\le F_B(k,t)$. Analogously, given $\kappa_L(x)$, function $g(x)$ must also
satisfy a similar realizability condition~\citep{moffatt78}. The later
results in the condition $-1\le h\le1$ in Eq.~(\ref{POWER_V}).
\label{REALIZABILITY}
}

To study a realistic case, we consider
velocity correlation tensor~(\ref{KAPPA_FOURIER}) with the Kolmogorov
power velocity spectrum. It is important to note that 
in the Kazantsev-Kraichnan model~(\ref{V_V_TENSOR}), the velocity field enters 
the eigenfunction equations~(\ref{quantum}, \ref{spinor}, \ref{W2_AND_W3}) only 
in the form of turbulent 
diffusivity $\kappa^{ij}(|{\bf x}-{\bf x}'|)=\int \langle v^i({\bf x}, t)v^j({\bf x}', t') \rangle d(t-t')$. 
In the Kolmogorov turbulence, the latter scales as $v_ll\sim l^{4/3}$, 
where $l=|{\bf x}-{\bf x}'|$ \citep[see, e.g.,][]{frisch}.  Comparing this with 
formula~(\ref{V_V_TENSOR}), we find that the Kolmogorov scaling implies 
$\kappa({\bf x})\approx\kappa(0)(x/l_0)^{4/3}\approx v_0 l_0(x/l_0)^{4/3}$. 
Without loss of generality we take $l_0\sim 1$, $v_0\sim 1$, and therefore 
\beq
\begin{array}{lcl}
F(k)&=&k^{-13/3},\\
G(k)&=&-hk^{-1}F(k),
\end{array}
\quad
2\le k\le k_{\rm max}.
\label{POWER_V}
\eeq
Here the minimal cutoff wavenumber $k_{\rm min}=2$, the maximal cutoff
wavenumber $k_{\rm max}\approx2[\kappa_L(0)/\nu]^{3/4}\approx4\nu^{-3/4}$ 
is determined by the plasma kinematic viscosity $\nu$, and the helicity
parameter $h$ must satisfy the realizability condition 
$-1\le h\le1$; the velocity field is
maximally helical when $|h|=1$. The resulting growth rates $\lambda_n$
of the bound (localized) eigenmodes, $\lambda_n>\lambda_0$, are
shown in Figure~\ref{FIGURE_1}, where the magnetic diffusivity is
chosen to be $\eta=10^{-6}$. 
The growth rates are mesured in units of large-scale eddy turnover 
rate~$\sim v_0/l_0$. 
The linear-scale plots~A and~B correspond to $h=1$ and $h=0.1$
respectively, while $k_{\rm max}=3$, which is consistent with the
Reynolds number being of order unity (i.e., a single-scale velocity field). 
The logarithmic-scale plots~C and~D correspond to $h=1$ and $h=0.1$,
while $k_{\rm max}=3000$, which is consistent with large
Reynolds and large magnetic Prandtl numbers.
Finally, the logarithmic-scale plots~E and~F correspond to $h=1$ and
$h=0.1$, while $k_{\rm max}=3\times10^7$, which is consistent with a
very large Reynolds number and a small Prandtl number. 

\section{Discussion and conclusion}

We find that when the Reynolds number is of order unity (a single-scale 
velocity field), the magnitude of the kinetic helicity parameter $h$ does 
not have much effect on the bound magnetic eigenmodes 
(plots~A and~B in Fig.~\ref{FIGURE_1}). 
However, in the case when the Reynolds number is large and the velocity
fluctuations extend over a large range of scales, the bound eigenmodes
are significantly affected by kinetic helicity at scales larger than
the viscous scale. 
(Magnetic fluctiations at the scales much smaller than the viscous scale, 
which are excited in the case ${\rm Pr}>>1$, are not significantly 
affected by magnetic helicity, which is consistent with previous 
considerations; e.g.,~\citet{kulsrud}.)
The number of bound eigenmodes increases
considerably when the kinetic helicity increases, and the corresponding 
eigenvalues, $\lambda_n$, become strongly concentrated near the 
eigenvalue of the fastest unbound eigenmode, $\lambda_0$. This result
follows from a nearly uniform distribution of $\lambda_n$ on the
logarithmic-scale plots in Fig.~\ref{FIGURE_1}.  

Moreover, we observe an important fact that in all the cases in 
Fig.~\ref{FIGURE_1} the growth rate of the first bound  
eigenmode,~$\lambda_1$, happens to be 
very close to the growth rate of the fastest
growing unbound eigenmode,~$\lambda_0$.\footnote{We found same result for all other 
high-Reynolds cases that we investigated (not reported here), with different helicity parameters.} 
We conjecture that 
for high Reynolds numbers the potential in~(\ref{W2_AND_W3}) always 
has a  shallow bound state~$\lambda_1$ 
(such that $\lambda_1 -\lambda_0\ll \lambda_0$)  
whose characteristic scale is much larger than the velocity 
correlation scale, due to~(\ref{W_LARGE_X_TRAPPED}). 
This has an important consequence for the dynamo mechanism. To understand it, we  
note that the alpha dynamo model~(\ref{alpha}), 
describing the large-scale magnetic field, does not capture the 
bound eigenmodes. However, our analysis shows that shallow bound 
eigenmodes have faster growth rates and large correlation 
lengths, so at a given scale~$x$ they may rapidly become 
dominant over the unbound modes. In practical applications, 
this means that such modes rather than the modes described by~(\ref{alpha})   
become essential in the large-scale magnetic field configurations. In this case  
the conventional alpha-dynamo model~(\ref{alpha}) leads to an inadequate  
description of the large-scale magnetic field 
even at the kinematic stage of dynamo action. 

We are grateful to Fausto Cattaneo and Alexander Obabko for many fruitful 
discussions. 
This work was supported by the NSF Center for Magnetic Self-Organization in 
Laboratory and Astrophysical Plasmas at the Universities of Chicago and 
Wisconsin--Madison, and by the US Department of Energy under grant 
DE-FG02-07ER54932.
We thank the Aspen Center for Physics, where a part 
of this work was done,  for hospitality and support.



\clearpage

\begin{figure}
\epsscale{1.0}
\plotone{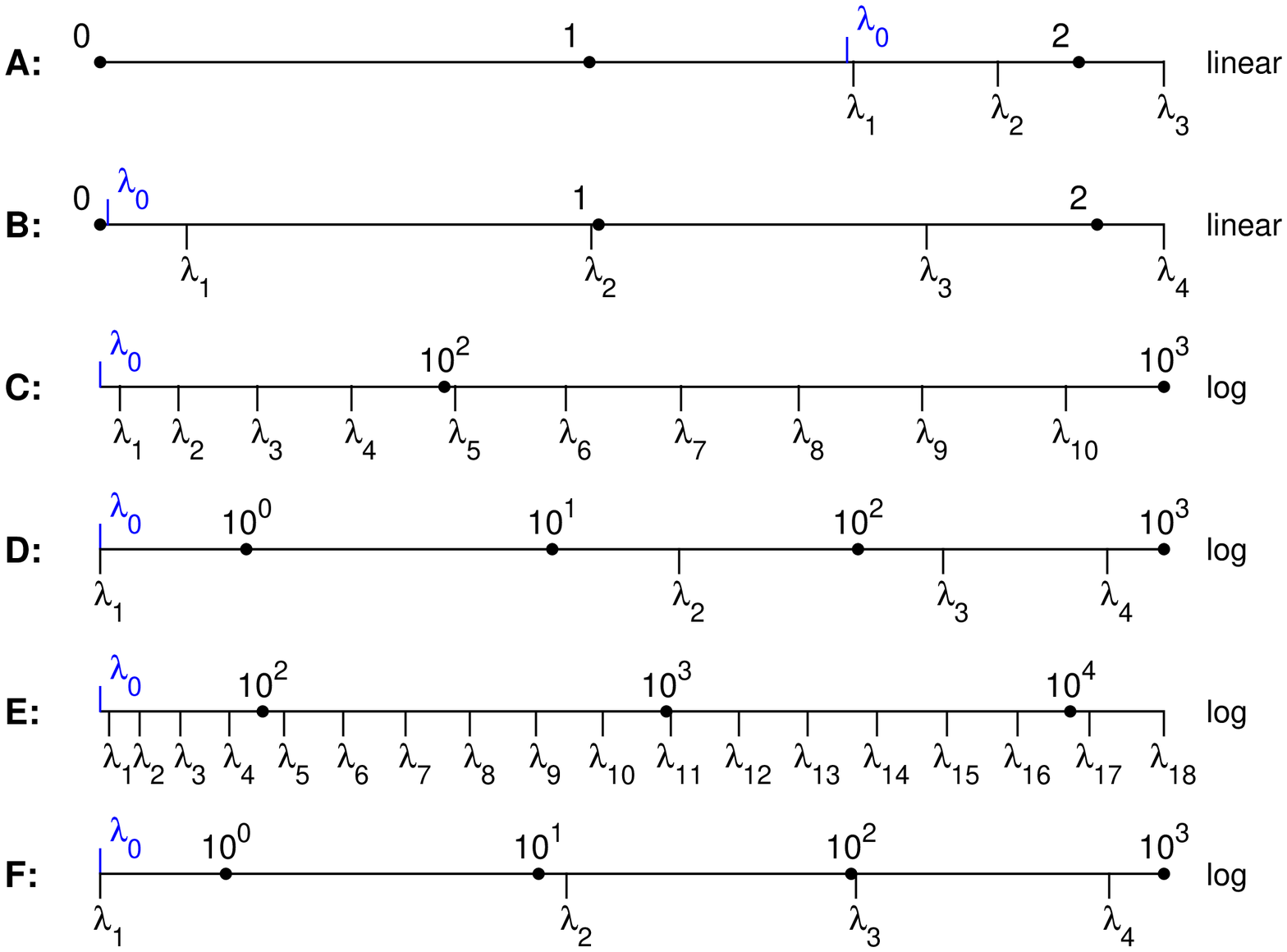}
\caption{Growth rates $\lambda_n > \lambda_0$ of the bound magnetic 
eigenmodes for the Kolmogorov velocity spectrum given by Eq.~(\ref{POWER_V}) 
and $\eta=10^{-6}$. Plots~A and~B are for $h=1$ and $0.1$, while 
$k_{\rm max}=3$ (Reynolds number ${\rm Re}\sim1$). Plots~C and~D are for
$h=1$ and $0.1$, while $k_{\rm max}=3000$ (${\rm Re}\gg1$ and Prandtl
number ${\rm Pr}\gg1$). Plots~E and~F are for $h=1$ and $0.1$, while
$k_{\rm max}=3\times10^7$ (${\rm Re}\gg1$ and ${\rm Pr}\ll1$).
\label{FIGURE_1}
}
\end{figure}

\end{document}